\documentclass[sigconf]{acmart}
\AtBeginDocument{%
  \providecommand\BibTeX{{%
    \normalfont B\kern-0.5em{\scshape i\kern-0.25em b}\kern-0.8em\TeX}}}

\usepackage[inline]{enumitem}
\usepackage{pbox}
\usepackage{balance}
\usepackage{todonotes}
\newlist{inlinelist}{enumerate*}{1}
\setlist*[inlinelist,1]{label=\roman*),itemjoin={{, }},itemjoin*={{, and }}}
\usepackage{url}
\usepackage[normalem]{ulem}
\usepackage{multirow}
\usepackage{multicol}
\usepackage{booktabs}

\begin{document}
\fancyhead{} 
\title{SPLADE v2: Sparse Lexical and Expansion Model for Information Retrieval}


\author{Thibault Formal}
\affiliation{%
  \institution{Naver Labs Europe}
  \city{Meylan}
  \country{France}
 }
 \affiliation{%
  \institution{Sorbonne Université, LIP6}
  \city{Paris}
  \country{France}
  }
  
\email{thibault.formal@naverlabs.com}

\author{Benjamin Piwowarski}
\affiliation{%
  \institution{Sorbonne Université, CNRS, LIP6}
  \city{Paris}
  \country{France}}
\email{benjamin.piwowarski@lip6.fr}

\author{Carlos Lassance}
\affiliation{%
  \institution{Naver Labs Europe}
  \city{Meylan}
  \country{France}}
\email{carlos.lassance@naverlabs.com}

\author{Stéphane Clinchant}
\affiliation{%
  \institution{Naver Labs Europe}
  \city{Meylan}
  \country{France}}
\email{stephane.clinchant@naverlabs.com}





\begin{abstract}
  
In neural Information Retrieval (IR), ongoing research is directed towards improving the first retriever in ranking pipelines. Learning dense embeddings to conduct retrieval using efficient approximate nearest neighbors methods has proven to work well. Meanwhile, there has been a growing interest in learning \emph{sparse} representations for documents and queries, that could inherit from the desirable properties of bag-of-words models such as the exact matching of terms and the efficiency of inverted indexes. Introduced recently, the SPLADE model provides highly sparse representations and competitive results with respect to state-of-the-art dense and sparse approaches. In this paper, we build on SPLADE and propose several significant improvements in terms of effectiveness and/or efficiency. More specifically, we modify the pooling mechanism, benchmark a model solely based on document expansion, and introduce models trained with distillation. We also report results on the BEIR benchmark. Overall, SPLADE is considerably improved with more than $9$\% gains on NDCG@10 on TREC DL 2019, leading to state-of-the-art results on the BEIR benchmark.

\end{abstract}






\settopmatter{printacmref=false} 

\maketitle

\section{Introduction}

The release of large pre-trained language models like BERT~\cite{bert} has shaken-up Natural Language Processing and Information Retrieval.
These models have shown a strong ability to adapt to various tasks by simple fine-tuning. At the beginning of 2019, \textit{Nogueira and Cho}~\cite{passage_ranking} achieved state-of-the-art results -- by a large margin -- on the MS MARCO passage re-ranking task, paving the way for LM-based neural ranking models. Because of strict efficiency requirements, these models have initially been used as re-rankers in a two-stage ranking pipeline, where first-stage retrieval -- or candidate generation -- is conducted with bag-of-words models (e.g. BM25) that rely on inverted indexes. While BOW models remain strong baselines~\cite{neural_hype}, they suffer from the long standing vocabulary mismatch problem, where relevant documents might not contain terms that appear in the query. Thus, there have been attempts to substitute standard BOW approaches by learned (neural) rankers. Designing such models poses several challenges regarding efficiency and scalability: therefore there is a need for methods where most of the computation can be done offline and online inference is fast. 
Dense retrieval with approximate nearest neighbors search has shown impressive results~\cite{xiong2021approximate,qu-etal-2021-rocketqa,lin-etal-2021-batch,Hofstaetter2021_tasb_dense_retrieval}, but can still benefit from BOW models (e.g. by combining both types of signals), due to the absence of explicit term matching. Hence, there has recently been a growing interest in learning \emph{sparse representations} for queries and documents~\cite{snrm, dai2019contextaware,nogueira2019document,zhao2020sparta,sparterm2020,gao-etal-2021-coil,10.1145/3404835.3463030,10.1145/3404835.3463098}. By doing so, models can inherit from the desirable properties of BOW models like exact-match of (possibly latent) terms, efficiency of inverted indexes and interpretability. Additionally, by modeling implicit or explicit (latent, contextualized) \emph{expansion} mechanisms -- similarly to standard expansion models in IR -- these models can reduce the vocabulary mismatch. 

In this paper, we build on the SPLADE model~\cite{10.1145/3404835.3463098} and propose several improvements/modifications that bring gains in terms of effectiveness or efficiency:
\begin{enumerate*}
\item by simply modifying SPLADE pooling mechanism, we are able to increase effectiveness by a large margin;
\item in the meantime, we propose an extension of the model without query expansion. Such model is inherently more efficient, as everything can be pre-computed and indexed offline, while providing results that are still competitive;
\item finally, we use distillation techniques~\cite{hofstatter2020improving} to boost SPLADE performance, leading to close to state-of-the-art results on the MS MARCO passage ranking task as well as the BEIR zero-shot evaluation benchmark~\cite{beir_2021}.
\end{enumerate*}



\section{Related Works}
Dense retrieval based on BERT Siamese models~\cite{sentence_bert} has become the standard approach for candidate generation in Question Answering and IR~\cite{guu2020realm,karpukhin2020dense,xiong2021approximate,qu-etal-2021-rocketqa,lin-etal-2021-batch,Hofstaetter2021_tasb_dense_retrieval}. 
While the backbone of these models remains the same, 
recent works highlight the critical aspects of the training strategy to obtain state-of-the-art results, ranging from improved negative sampling~\cite{lin-etal-2021-batch,Hofstaetter2021_tasb_dense_retrieval} to distillation~\cite{hofstatter2020improving,lin-etal-2021-batch}. ColBERT~\cite{colbert} pushes things further: the postponed token-level interactions allow to efficiently apply the model for first-stage retrieval, benefiting of the effectiveness of modeling fine-grained interactions, at the cost of storing embeddings for each (sub)term -- raising concerns about the actual scalability of the approach for large collections. 
To the best of our knowledge, very few studies have discussed the impact of using \emph{approximate} nearest neighbors (ANN) search on IR metrics~\cite{boytsov2018efficient, tu2020approximate}.
Due to the moderate size of the MS MARCO collection, results are usually reported  with an \emph{exact}, brute-force search, therefore giving no indication on the effective computing cost.

An alternative to dense indexes is term-based ones.
Building on standard BOW models, \textit{Zamani et al.} first introduced SNRM~\cite{snrm}: the model embeds documents and queries in a sparse high-dimensional latent space by means of $\ell_1$ regularization on representations.
However, SNRM effectiveness remains limited and its efficiency has been questioned~\cite{paria2020minimizing}. 

Motivated by the success of BERT, there have been attempts to transfer the knowledge from pre-trained LM to sparse approaches. DeepCT~\cite{dai2019contextaware, 10.1145/3366423.3380258, 10.1145/3397271.3401204} focused on learning contextualized term weights in the full vocabulary space -- akin to BOW term weights. 
However, as the vocabulary associated with a document remains the same, this type of approach does not solve the vocabulary mismatch, as acknowledged by the use of query expansion for retrieval \cite{dai2019contextaware}.
A first solution to this problem consists in expanding documents using generative approaches such as doc2query~\cite{nogueira2019document} and doc2query-T5~\cite{doct5} to predict expansion words for documents. The document expansion adds new terms to documents -- hence fighting the vocabulary mismatch -- as well as repeats existing terms, implicitly performing re-weighting by boosting important terms. 

Recently, DeepImpact~\cite{10.1145/3404835.3463030} combined the expansion from doc2query-T5 with the re-weighting from DeepCT to learn term \emph{impacts}. These expansion techniques are however limited by the way they are trained (predicting queries), which is indirect in nature and limit their progress.
A second solution to this problem, that has been chosen by recent works~\cite{sparterm2020,MacAvaney_2020,zhao2020sparta,10.1145/3404835.3463098}, is to estimate the importance of each term of the vocabulary \emph{implied by} each term of the document (or query), i.e. to compute an interaction matrix between the document or query tokens and all the tokens from the vocabulary. This is followed by an aggregation mechanism (roughly sum for SparTerm~\cite{sparterm2020} and SPLADE~\cite{10.1145/3404835.3463098}, max for EPIC~\cite{MacAvaney_2020} and SPARTA~\cite{zhao2020sparta}), that allows to compute an importance weight for each term of the vocabulary, for the full document or query. 

However, EPIC and SPARTA (document) representations are not sparse enough by construction -- unless resorting on top-$k$ pooling -- contrary to SparTerm, for which fast retrieval is thus possible. Furthermore, the latter does not include (like SNRM) an \emph{explicit} sparsity regularization, which hinders its performance. SPLADE however relies on such regularization, as well as other key changes, that boost both the efficiency and the effectiveness of this type of approaches, providing a model that both learns expansion and compression in an end-to-end manner.
Furthermore, COIL~\cite{gao-etal-2021-coil} proposed to revisit exact-match mechanisms by learning dense representations \emph{per term} to perform contextualized term matching, at the cost of increased index size.

\section{Sparse Lexical representations for first-stage ranking}


In this section, we first describe in details the SPLADE model recently introduced in \cite{10.1145/3404835.3463098}.

\subsection{SPLADE}

SPLADE  predicts term importance -- in BERT WordPiece vocabulary ($|V|=30522$) -- based on the logits of the Masked Language Model (MLM) layer. More precisely, let us consider an input query or document sequence (after WordPiece tokenization) $t=(t_1,t_2,...,t_N)$, and its corresponding BERT embeddings $(h_1,h_2,...,h_N)$. We consider the importance $w_{ij}$ of the token $j$ (vocabulary) for a token $i$ (of the input sequence): 
\begin{equation}
    w_{ij} = \text{transform}(h_i)^T E_j + b_j \quad j \in \{1,...,|V|\}
    \label{equation_1}
\end{equation} 
where $E_j$ denotes the BERT input embedding for token $j$, $b_j$ is a token-level bias, and transform$(.)$ is a linear layer with GeLU activation and LayerNorm. Note that Eq.~\eqref{equation_1} is equivalent to the MLM prediction, thus it can also be initialized from a pre-trained MLM model. The final representation is then obtained by summing importance predictors over the input sequence tokens, after applying a log-saturation effect~\cite{10.1145/3404835.3463098}: 
\begin{equation}\label{eq:2}
w_j= \sum_{i \in t} \log \left(1 + \text{ReLU}(w_{ij}) \right)
\end{equation}

\paragraph{\bf Ranking loss} 
\noindent Let $s(q,d)$ denote the ranking score obtained via dot product between $q$ and $d$ representations from Eq.~\eqref{eq:2}. Given a query $q_i$ in a batch, a positive document $d_i^+$, a (hard) negative document $d_i^-$ (e.g. coming from BM25 sampling), and a set of negative documents in the batch (positive documents from other queries) $\{d_{i,j}^-\}_j$, we consider a constrastive loss, which can be interpreted as the maximization of the probability of the document $d_i^+$ being relevant among the documents $d_i^+, d_i^-$ and $\{d_{i,j}^-\}$:

\begin{equation}\mathcal{L}_{rank-IBN} = - \log\frac{e^{s(q_i,d_i^+)}}{e^{s(q_i,d_i^+)} + e^{s(q_i,d_i^-)} + \sum_j e^{s(q_i,d_{i,j}^-)}}
\end{equation}

The \emph{in-batch negatives} (IBN) sampling strategy is widely used for training image retrieval models, and has shown to be effective in learning first-stage rankers~\cite{karpukhin2020dense,qu-etal-2021-rocketqa,lin-etal-2021-batch}.

\paragraph{\bf Learning sparse representations}
The idea of learning sparse representations for first-stage retrieval dates back to SNRM~\cite{snrm}, via $\ell_1$ regularization. Later, \cite{paria2020minimizing} pointed-out that minimizing the $\ell_1$ norm of representations does not result in the most efficient index, as nothing ensures that posting lists are evenly distributed. Note that this is even more true for standard indexes due to the Zipfian nature of the term frequency distribution. 
To obtain a well-balanced index, \textit{Paria et al.}~\cite{paria2020minimizing} introduce the \texttt{FLOPS} regularizer, a smooth relaxation of the average number of floating-point operations necessary to compute the score between a query and a document, and hence directly related to the retrieval time. It is defined using $a_j$ as a continuous relaxation of the activation (i.e. the term has a non-zero weight) probability $p_j$ for token $j$, and estimated for documents $d$ in a batch of size $N$ by
$\bar{a}_j=\frac{1}{N} \sum_{i=1}^N w^{(d_i)}_{j}$.
This gives the following regularization loss
\begin{equation}\label{eq:reg}
\ell_{\texttt{FLOPS}} = \sum_{j\in V} {\bar a}_j^2 = \sum_{j \in V} \left( \frac{1}{N} \sum_{i=1}^N  w_j^{(d_i)} \right)^2
\end{equation}
%

\paragraph{\bf Overall loss}
By jointly optimizing the model in Eq.~\eqref{eq:2} with ranking and regularization losses, SPLADE combines the best of both worlds for end-to-end training of sparse, expansion-aware representations of documents and queries:
\begin{equation}
\mathcal{L} = \mathcal{L}_{rank-IBN} + \lambda_q \mathcal{L}^{q}_{\texttt{reg}} + \lambda_d \mathcal{L}^{d}_{\texttt{reg}}
\end{equation} 
where $\mathcal{L}_{\texttt{reg}}$ is the sparse \texttt{FLOPS} regularization from Eq.~\ref{eq:reg}. We use two distinct regularization weights ($\lambda_d$ and $\lambda_q$) for queries and documents -- allowing to put more pressure on the sparsity for queries, which is critical for fast retrieval.

\subsection{Pooling strategy}

We propose to change the sum in Eq.~\eqref{eq:2} by a $\max$ pooling operation:
\begin{equation}
w_j= \max_{i \in t} \log \left(1 + \text{ReLU}(w_{ij}) \right)
\end{equation}
The model now bears more similarities with SPARTA and EPIC, and to some extent ColBERT. As shown in the experiments section, it considerably improves SPLADE performance. In the following, max pooling is the default configuration for SPLADE, and the corresponding model is referred to as SPLADE-${\max}$.

\subsection{SPLADE document encoder}
In addition to the max pooling operation, we consider a document-only version of SPLADE.
In this case, there are no query expansion nor query term weighting, and the ranking score is simply given by:
\begin{equation}
 s(q,d) = \sum_{j  \in q}  w_j^d
\end{equation}

Such extension offers an interesting efficiency boost: because the ranking score solely depends on the document term weights, everything can be pre-computed offline, and inference cost is consequently reduced, while still offering competitive results as shown in the experiments. We refer to this model as SPLADE-doc.

\subsection{Distillation and hard negatives}
We also incorporate distillation to our training procedure, following the improvements shown in~\cite{hofstatter2020improving}. The distillation training is done in two steps: \begin{enumerate*} \item we first train both a SPLADE first-stage retriever as well as a cross-encoder reranker~\footnote{Using \url{https://huggingface.co/cross-encoder/ms-marco-MiniLM-L-12-v2} as pre-trained checkpoint.} using the triplets generated by~\cite{hofstatter2020improving}; \item in the second step, we generate triplets using SPLADE trained with distillation (thus providing harder negatives than BM25), and use the aforementioned reranker to generate the scores needed for the Margin-MSE loss. We then train a SPLADE model from scratch using these triplets and scores. The result of the second step is what we call DistilSPLADE-${\max}$. 
\end{enumerate*}

\section{Experimental setting and results}

We trained and evaluated our models on the MS MARCO passage ranking dataset\footnote{\url{https://github.com/microsoft/MSMARCO-Passage-Ranking}} in the full ranking setting. The dataset contains approximately $8.8$M passages, and hundreds of thousands training queries with shallow annotation ($\approx 1.1$ relevant passages per query in average). The development set contains $6980$ queries with similar labels, while the TREC DL 2019 evaluation set provides fine-grained annotations from human assessors for a set of $43$ queries~\cite{craswell2020overview}. 

\paragraph{\bf Training, indexing and retrieval}

We initialized the models with the \texttt{DistilBERT-base} checkpoint. Models are trained with the ADAM optimizer, using a learning rate of $2e^{-5}$ with linear scheduling and a warmup of $6000$ steps, and a batch size of $124$. We keep the best checkpoint using MRR@10 on a validation set of $500$ queries, after training for $150$k iterations, using an approximate retrieval validation set similar to~\cite{Hofstaetter2021_tasb_dense_retrieval}. For the SPLADE-doc approach, we simply train for $50$k steps and select the last checkpoint. 
We consider a maximum length of $256$ for input sequences. In order to mitigate the contribution of the regularizer at the early stages of training, we follow \cite{paria2020minimizing} and use a scheduler for $\lambda$, quadratically increasing $\lambda$ at each training iteration, until a given step ($50$k in our case), from which it remains constant. Typical values for $\lambda$ fall between $1e^{-1}$ and $1e^{-4}$.
For storing the index, 
we use a custom implementation based on Python arrays, and we rely on Numba~\cite{lam2015numba} to parallelize retrieval. Models\footnote{We made the code public at \url{https://github.com/naver/splade}} are trained using PyTorch~\cite{paszke2019pytorch} and HuggingFace transformers~\cite{wolf2020huggingfaces}, on $4$ Tesla $V100$ GPUs with $32$GB memory. 

\paragraph{\bf Evaluation}
We report Recall@1000 for both datasets, as well as the official metrics MRR@10 and NDCG@10 for MS MARCO dev set and TREC DL 2019 respectively. Since we are essentially interested in the first retrieval step,  we do not consider re-rankers based on BERT, and we compare our approach to first stage rankers only -- results reported on the MS MARCO leaderboard are thus not comparable to the results presented here. We compare to the following sparse approaches
\begin{enumerate*}
    \item BM25,
    \item DeepCT~\cite{dai2019contextaware}, 
    \item doc2query-T5~\cite{doct5},
    \item SparTerm~\cite{sparterm2020},
    \item COIL-tok~\cite{gao-etal-2021-coil} and
    \item DeepImpact~\cite{10.1145/3404835.3463030}
\end{enumerate*}, as well as state-of-the-art dense approaches ANCE~\cite{xiong2020approximate}, TCT-ColBERT~\cite{lin-etal-2021-batch} and TAS-B~\cite{Hofstaetter2021_tasb_dense_retrieval}, reporting results from corresponding papers.

\begin{figure}[b]
  \caption{Performance vs FLOPS for SPLADE models trained with different regularization strength $\lambda$ on MS MARCO.}
  \includegraphics[width=0.45\textwidth]{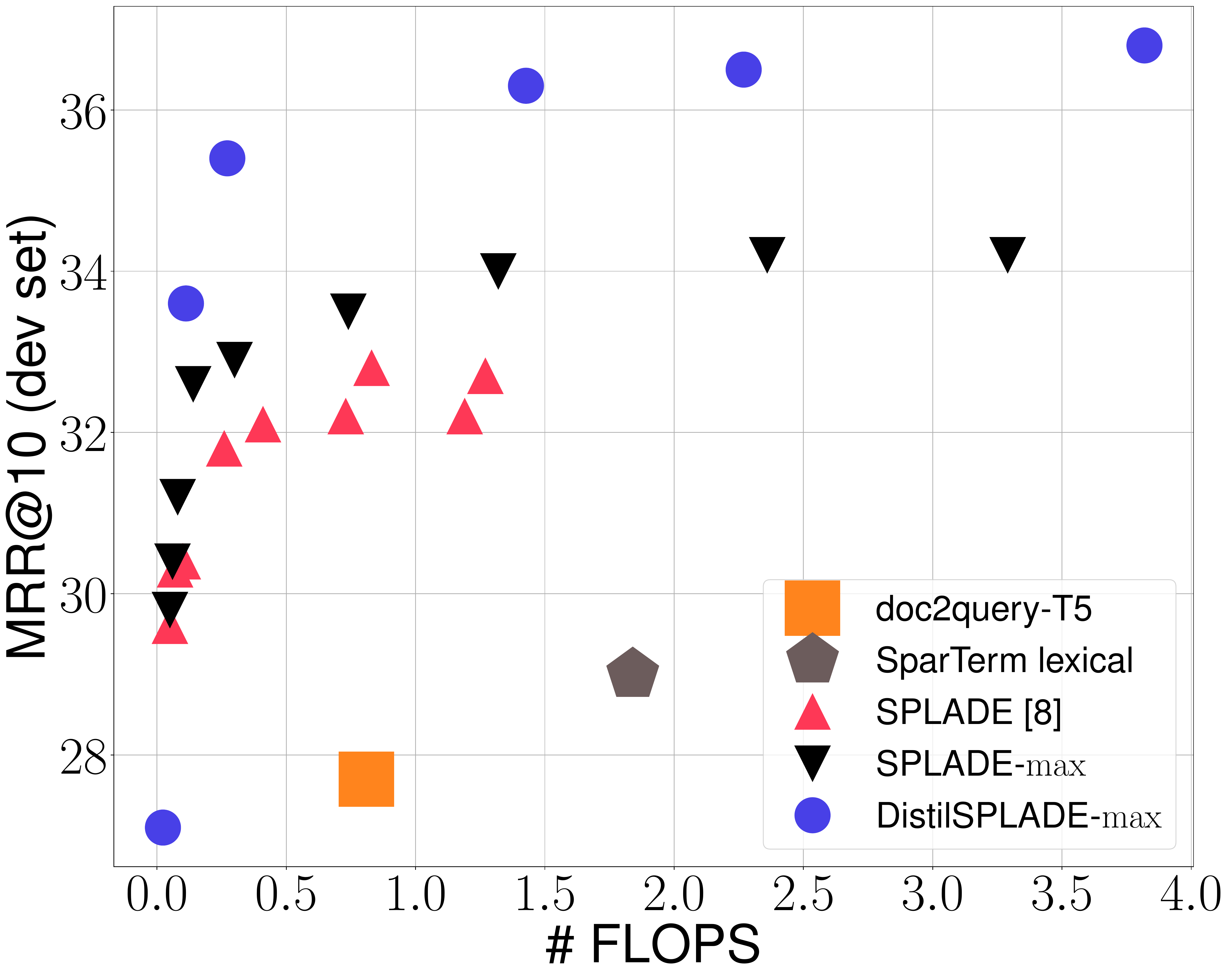}
  \label{perf_flops}
\end{figure}

MS MARCO dev and TREC DL 2019 results are given in Table \ref{table_1}: as the performance for SPLADE depends on the regularization strength $\lambda$, and as more efficient models are generally less effective, we select in the table the best performing model in our grid of experiments, with reasonable efficiency (in terms of FLOPS). Figure \ref{perf_flops} highlights the actual trade-off between effectiveness and efficiency for SPLADE, by showing the performance (MRR@10 on MS MARCO dev set) vs FLOPS, for SPLADE models trained with different regularization strength. The FLOPS metric is an estimation of the average number of floating-point operations between a query and a document which is defined as 
the expectation $\mathbb{E}_{q,d} \left[ \sum_{j \in V} p_j^{(q)}p_j^{(d)} \right]$
where $p_j$ is the activation probability for token $j$ in a document $d$ or a query $q$. It is empirically estimated from a set of approximately $100$k development queries, on the MS MARCO collection. Overall, we observe that:
\begin{enumerate*}
    \item \textit{our improved models outperform the other sparse retrieval methods by a large margin on the MS MARCO dev set as well as TREC DL 2019 queries};
    \item \textit{the results are competitive with state-of-the-art dense retrieval methods.} 
\end{enumerate*}

\paragraph{\bf BEIR}
Finally, we verify the zero-shot performance of SPLADE using a subset of datasets from the BEIR~\cite{beir_2021} benchmark that encompasses various IR datasets for zero-shot evaluation. We solely use a subset due to the fact that some of the datasets (namely \texttt{CQADupstack}, \texttt{BioASQ}, \texttt{Signal-1M}, \texttt{TREC-NEWS},	\texttt{Robust04}) are not readily available. Results are displayed in Table~\ref{table:beir_ndcg} (NDCG@10). We compare against the best performing models from the original benchmark paper~\cite{beir_2021} (ColBERT~\cite{colbert}) and the two best performing from the rolling benchmark\footnote{\url{https://docs.google.com/spreadsheets/d/1L8aACyPaXrL8iEelJLGqlMqXKPX2oSP_R10pZoy77Ns}} (tuned BM25 and TAS-B~\cite{Hofstaetter2021_tasb_dense_retrieval}). We also report the SPLADE evaluation against these baselines.

\setlength{\tabcolsep}{2pt} 
\begin{table}
\centering
\caption{Evaluation on MS MARCO passage retrieval (dev set) and TREC DL 2019.}
\begin{tabular}{lcccc}
\toprule
model &  \multicolumn{2}{c}{MS MARCO dev} & \multicolumn{2}{c}{TREC DL 2019} \\
& MRR@10 & R@1000 & NDCG@10 & R@1000  \\
\midrule
\texttt{Dense retrieval} & & &  \\
Siamese (ours) & 0.312 & 0.941 & 0.637 & 0.711   \\
ANCE~\cite{xiong2020approximate} & 0.330 & 0.959 & 0.648 & -   \\
TCT-ColBERT~\cite{lin-etal-2021-batch} & 0.359 & 0.970 & 0.719 & 0.760  \\
TAS-B~\cite{Hofstaetter2021_tasb_dense_retrieval}       &  0.347 &  0.978 & 0.717 & 0.843  \\
RocketQA~\cite{qu-etal-2021-rocketqa} &  0.370 &  0.979 & - & -  \\
\bottomrule
\hline
\texttt{Sparse retrieval} & & & & \\
BM25 & 0.184 & 0.853 & 0.506 & 0.745  \\ 
DeepCT~\cite{dai2019contextaware} & 0.243 & 0.913 & 0.551 & 0.756  \\ 
doc2query-T5~\cite{doct5} & 0.277 & 0.947 & 0.642 & 0.827\\
SparTerm~\cite{sparterm2020} & 0.279 & 0.925 & - & - \\
COIL-tok~\cite{gao-etal-2021-coil}                         &     0.341 &  0.949 &   0.660 & -   \\
DeepImpact~\cite{10.1145/3404835.3463030}                         &     0.326 & 0.948 & 0.695 &  -  \\
SPLADE~\cite{10.1145/3404835.3463098}  & 0.322   & 0.955  & 0.665 & 0.813   \\
\bottomrule
\hline 
\texttt{Our methods} & & & &  \\

SPLADE-${\max}$               & 0.340  &  0.965 & 0.684  &  0.851           \\
SPLADE-doc                       &  0.322	& 	0.946	&  0.667 &  0.747  \\
DistilSPLADE-$\max$           &  0.368	& 	0.979	&  0.729 & 0.865   \\

\end{tabular}
\label{table_1}
\end{table}

\begin{figure}[b]
  \caption{Performance vs average document length (number of non-zero dimensions in document representations) for SPLADE-doc models trained with different regularization strength $\lambda_d$ on MS MARCO.}
  \includegraphics[width=0.45\textwidth]{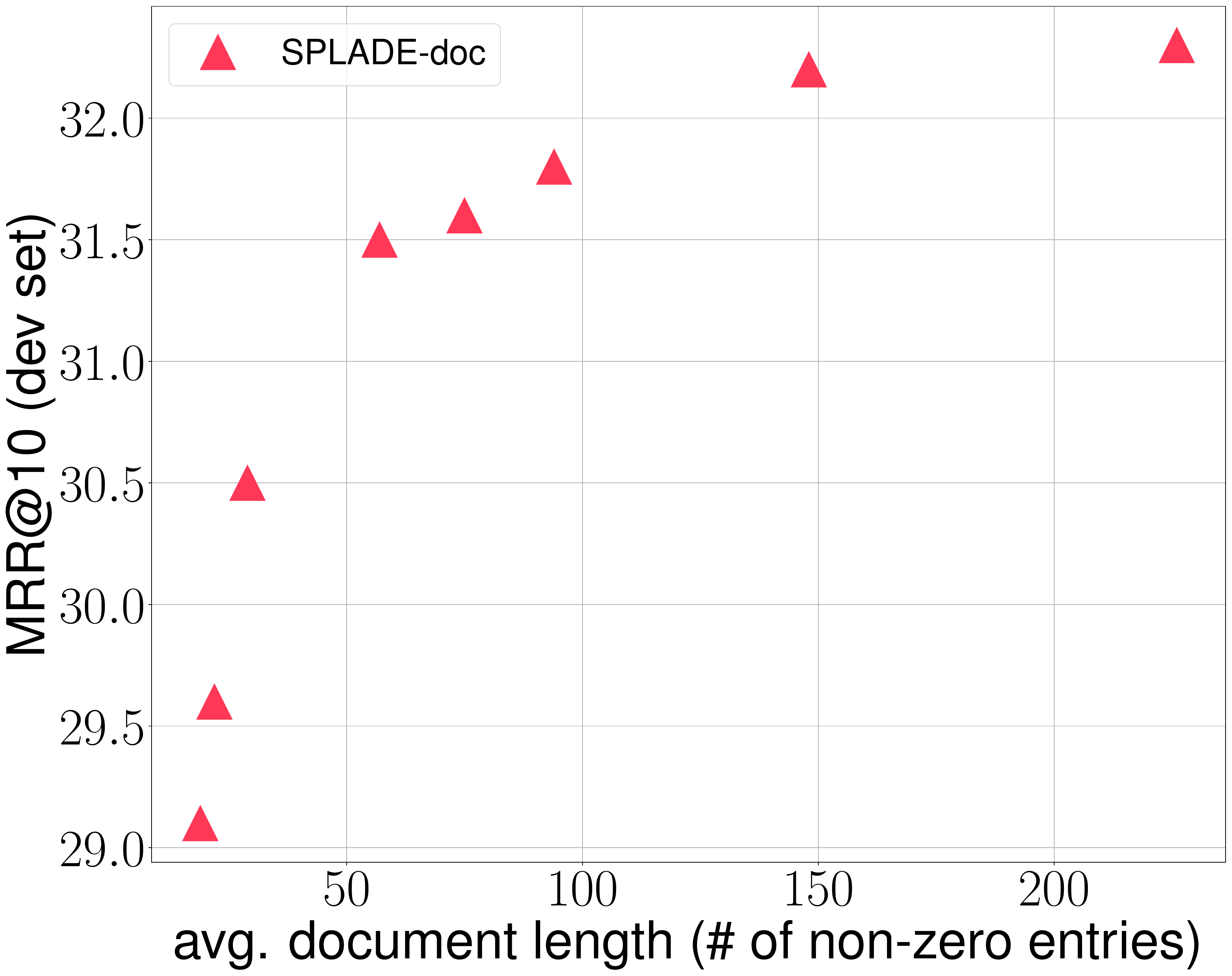}
  \label{splade_doc}
\end{figure}


\subsection{Impact of max pooling}

First, on MS MARCO and TREC, max pooling brings almost $2$ points in MRR@10 and NDCG@10 compared to the SPLADE baseline.
It becomes competitive with COIL and DeepImpact. In addition, Figure~\ref{perf_flops} shows that SPLADE-$\max$ is consistently better than SPLADE,
in terms of effectiveness-efficiency trade-off. SPLADE-$\max$ has also improved performance on the BEIR benchmark
(cf Table \ref{table:beir_ndcg}).

\subsection{Document expansion}

Our document encoder model with max pooling is able to reach the same performance as the previous SPLADE model, outperforming doc2query-T5 on MS MARCO. As this model has no query encoder, it is more efficient in terms of e.g. latency. Figure~\ref{splade_doc} illustrates how we can balance efficiency (in terms of the average size of document representations) with effectiveness. For relatively sparse representations, we are able to obtain performance on par with approaches like doc2query-T5 (e.g. MRR@10=$29.6$ for a model with an average of $19$ non-zero weights per document).
In addition, it is straightforward to train and apply to a new document collection: a single forward is required as opposed to multiple inferences with beam search for doc2query-T5. 

\subsection{Distillation}

By training with distillation, we are able to considerably improve the performance of SPLADE, as seen in Table~\ref{table_1}. From Figure~\ref{perf_flops}, we observe that distilled models bring huge improvements for higher values of FLOPS ($0.368$ MRR@10 for $\approx$ $4$ FLOPS), but are still very efficient in low regime ($0.35$ MRR for $\approx$ 0.3 FLOPS).
Furthermore, DistilSPLADE-$\max$ is able to outperform all other methods in most datasets of the BEIR benchmark (cf Table~\ref{table:beir_ndcg}). 

\begin{table}[]
\caption{NDCG@10 results on BEIR (subset containing all the readily available datasets).}
\label{table:beir_ndcg}
\begin{tabular}{@{}lccc|ccc@{}}
\toprule
\multirow{2}{*}{Corpus} & \multicolumn{3}{c}{Baselines}           & \multicolumn{3}{|c}{SPLADE}    \\ \cmidrule(l){2-7} 
                        & ColBERT        & BM25 & TAS-B & sum~\cite{10.1145/3404835.3463098}   & max   & distil   \\ \midrule
\texttt{MS MARCO}                 & 0.425          & 0.228           & 0.408 & 0.387 & 0.402 & \textbf{0.433} \\ 
\texttt{ArguAna}                 & 0.233          & 0.315           & 0.427 & 0.447 & 0.439 & \textbf{0.479} \\ 
\texttt{Climate-FEVER}           & 0.184          & 0.213           & 0.228 & 0.162 & 0.199 & \textbf{0.235} \\ 
\texttt{DBPedia}                 & 0.392          & 0.273           & 0.384 & 0.343 & 0.366 & \textbf{0.435} \\ 
\texttt{FEVER}                   & 0.771          & 0.753           & 0.700 & 0.728 & 0.730 & \textbf{0.786} \\ 
\texttt{FiQA-2018}                    & 0.317          & 0.236           & 0.300 & 0.258 & 0.287 & \textbf{0.336} \\ 
\texttt{HotpotQA}                & 0.593          & 0.603           & 0.584 & 0.635 & 0.636 & \textbf{0.684} \\ 
\texttt{NFCorpus}                & 0.305          & 0.325           & 0.319 & 0.311 & 0.313 & \textbf{0.334} \\ 
\texttt{NQ}                      & \textbf{0.524} & 0.329           & 0.463 & 0.438 & 0.469 & 0.521          \\ 
\texttt{Quora}                   & \textbf{0.854} & 0.789           & 0.835 & 0.829 & 0.835 & 0.838          \\ 
\texttt{SCIDOCS}                 & 0.145          & \textbf{0.158}  & 0.149 & 0.141 & 0.145 & \textbf{0.158} \\ 
\texttt{SciFact}                 & 0.671          & 0.665           & 0.643 & 0.626 & 0.628 & \textbf{0.693} \\ 
\texttt{TREC-COVID}              & 0.677          & 0.656           & 0.481 & 0.655 & 0.673 & \textbf{0.710} \\ 
\texttt{Touché-2020 (v1)}        & 0.275          & \textbf{0.614}  & 0.173 & 0.289 & 0.316 & 0.364          \\ \midrule
Avg. all             & 0.455          & 0.440           & 0.435 & 0.446 & 0.460 & \textbf{0.500}          \\ 
Avg. zero-shot       & 0.457          & 0.456           & 0.437 & 0.451 & 0.464 & \textbf{0.506}          \\ 
Best on dataset         & 2              & 2               & 0     & 0     & 0     & \textbf{11}             \\ \bottomrule
\end{tabular}
\end{table}

\section{Conclusion}
In this paper, we have built on the SPLADE model by reconsidering its pooling mechanism, and by using standard training techniques such as distillation for neural IR models. Our experiments have shown that the max pooling technique indeed provides a substantial improvement. Secondly,
the document encoder is an interesting model for faster retrieval conditions. Finally, the distilled SPLADE model leads to close to state-of-the-art models on MS MARCO and TREC DL 2019, while clearly outperforming recent dense models on zero-shot evaluation.



\bibliographystyle{ACM-Reference-Format}
\balance
\bibliography{sample-base}


\begin{thebibliography}{33}


\ifx \showCODEN    \undefined \def \showCODEN     #1{\unskip}     \fi
\ifx \showDOI      \undefined \def \showDOI       #1{#1}\fi
\ifx \showISBNx    \undefined \def \showISBNx     #1{\unskip}     \fi
\ifx \showISBNxiii \undefined \def \showISBNxiii  #1{\unskip}     \fi
\ifx \showISSN     \undefined \def \showISSN      #1{\unskip}     \fi
\ifx \showLCCN     \undefined \def \showLCCN      #1{\unskip}     \fi
\ifx \shownote     \undefined \def \shownote      #1{#1}          \fi
\ifx \showarticletitle \undefined \def \showarticletitle #1{#1}   \fi
\ifx \showURL      \undefined \def \showURL       {\relax}        \fi
\providecommand\bibfield[2]{#2}
\providecommand\bibinfo[2]{#2}
\providecommand\natexlab[1]{#1}
\providecommand\showeprint[2][]{arXiv:#2}

\bibitem[\protect\citeauthoryear{Bai, Li, Wang, Zhang, Shang, Xu, Wang, Wang,
  and Liu}{Bai et~al\mbox{.}}{2020}]%
        {sparterm2020}
\bibfield{author}{\bibinfo{person}{Yang Bai}, \bibinfo{person}{Xiaoguang Li},
  \bibinfo{person}{Gang Wang}, \bibinfo{person}{Chaoliang Zhang},
  \bibinfo{person}{Lifeng Shang}, \bibinfo{person}{Jun Xu},
  \bibinfo{person}{Zhaowei Wang}, \bibinfo{person}{Fangshan Wang}, {and}
  \bibinfo{person}{Qun Liu}.} \bibinfo{year}{2020}\natexlab{}.
\newblock \bibinfo{title}{SparTerm: Learning Term-based Sparse Representation
  for Fast Text Retrieval}.
\newblock
\newblock
\showeprint[arxiv]{2010.00768}~[cs.IR]


\bibitem[\protect\citeauthoryear{Boytsov}{Boytsov}{2018}]%
        {boytsov2018efficient}
\bibfield{author}{\bibinfo{person}{Leonid Boytsov}.}
  \bibinfo{year}{2018}\natexlab{}.
\newblock \emph{\bibinfo{title}{Efficient and Accurate Non-Metric k-NN Search
  with Applications to Text Matching}}.
\newblock \bibinfo{thesistype}{Ph.D. Dissertation}. \bibinfo{school}{Carnegie
  Mellon University}.
\newblock


\bibitem[\protect\citeauthoryear{Craswell, Mitra, Yilmaz, Campos, and
  Voorhees}{Craswell et~al\mbox{.}}{2020}]%
        {craswell2020overview}
\bibfield{author}{\bibinfo{person}{Nick Craswell}, \bibinfo{person}{Bhaskar
  Mitra}, \bibinfo{person}{Emine Yilmaz}, \bibinfo{person}{Daniel Campos},
  {and} \bibinfo{person}{Ellen~M Voorhees}.} \bibinfo{year}{2020}\natexlab{}.
\newblock \showarticletitle{Overview of the trec 2019 deep learning track}.
\newblock \bibinfo{journal}{\emph{arXiv preprint arXiv:2003.07820}}
  (\bibinfo{year}{2020}).
\newblock


\bibitem[\protect\citeauthoryear{Dai and Callan}{Dai and Callan}{2019}]%
        {dai2019contextaware}
\bibfield{author}{\bibinfo{person}{Zhuyun Dai} {and} \bibinfo{person}{Jamie
  Callan}.} \bibinfo{year}{2019}\natexlab{}.
\newblock \bibinfo{title}{Context-Aware Sentence/Passage Term Importance
  Estimation For First Stage Retrieval}.
\newblock
\newblock
\showeprint[arxiv]{1910.10687}~[cs.IR]


\bibitem[\protect\citeauthoryear{Dai and Callan}{Dai and Callan}{2020a}]%
        {10.1145/3366423.3380258}
\bibfield{author}{\bibinfo{person}{Zhuyun Dai} {and} \bibinfo{person}{Jamie
  Callan}.} \bibinfo{year}{2020}\natexlab{a}.
\newblock \bibinfo{booktitle}{\emph{Context-Aware Document Term Weighting for
  Ad-Hoc Search}}.
\newblock \bibinfo{publisher}{Association for Computing Machinery},
  \bibinfo{address}{New York, NY, USA}, \bibinfo{pages}{1897–1907}.
\newblock
\showISBNx{9781450370233}
\urldef\tempurl%
\url{https://doi.org/10.1145/3366423.3380258}
\showURL{%
\tempurl}


\bibitem[\protect\citeauthoryear{Dai and Callan}{Dai and Callan}{2020b}]%
        {10.1145/3397271.3401204}
\bibfield{author}{\bibinfo{person}{Zhuyun Dai} {and} \bibinfo{person}{Jamie
  Callan}.} \bibinfo{year}{2020}\natexlab{b}.
\newblock \bibinfo{booktitle}{\emph{Context-Aware Term Weighting For First
  Stage Passage Retrieval}}.
\newblock \bibinfo{publisher}{Association for Computing Machinery},
  \bibinfo{address}{New York, NY, USA}, \bibinfo{pages}{1533–1536}.
\newblock
\showISBNx{9781450380164}
\urldef\tempurl%
\url{https://doi.org/10.1145/3397271.3401204}
\showURL{%
\tempurl}


\bibitem[\protect\citeauthoryear{Devlin, Chang, Lee, and Toutanova}{Devlin
  et~al\mbox{.}}{2018}]%
        {bert}
\bibfield{author}{\bibinfo{person}{Jacob Devlin}, \bibinfo{person}{Ming{-}Wei
  Chang}, \bibinfo{person}{Kenton Lee}, {and} \bibinfo{person}{Kristina
  Toutanova}.} \bibinfo{year}{2018}\natexlab{}.
\newblock \showarticletitle{{BERT:} Pre-training of Deep Bidirectional
  Transformers for Language Understanding}.
\newblock \bibinfo{journal}{\emph{CoRR}}  \bibinfo{volume}{abs/1810.04805}
  (\bibinfo{year}{2018}).
\newblock
\showeprint[arxiv]{1810.04805}
\urldef\tempurl%
\url{http://arxiv.org/abs/1810.04805}
\showURL{%
\tempurl}


\bibitem[\protect\citeauthoryear{Formal, Piwowarski, and Clinchant}{Formal
  et~al\mbox{.}}{2021}]%
        {10.1145/3404835.3463098}
\bibfield{author}{\bibinfo{person}{Thibault Formal}, \bibinfo{person}{Benjamin
  Piwowarski}, {and} \bibinfo{person}{St\'{e}phane Clinchant}.}
  \bibinfo{year}{2021}\natexlab{}.
\newblock \showarticletitle{SPLADE: Sparse Lexical and Expansion Model for
  First Stage Ranking}. In \bibinfo{booktitle}{\emph{Proceedings of the 44th
  International ACM SIGIR Conference on Research and Development in Information
  Retrieval}} (Virtual Event, Canada) \emph{(\bibinfo{series}{SIGIR '21})}.
  \bibinfo{publisher}{Association for Computing Machinery},
  \bibinfo{address}{New York, NY, USA}, \bibinfo{pages}{2288–2292}.
\newblock
\showISBNx{9781450380379}
\urldef\tempurl%
\url{https://doi.org/10.1145/3404835.3463098}
\showDOI{\tempurl}


\bibitem[\protect\citeauthoryear{Gao, Dai, and Callan}{Gao
  et~al\mbox{.}}{2021}]%
        {gao-etal-2021-coil}
\bibfield{author}{\bibinfo{person}{Luyu Gao}, \bibinfo{person}{Zhuyun Dai},
  {and} \bibinfo{person}{Jamie Callan}.} \bibinfo{year}{2021}\natexlab{}.
\newblock \showarticletitle{{COIL}: Revisit Exact Lexical Match in Information
  Retrieval with Contextualized Inverted List}. In
  \bibinfo{booktitle}{\emph{Proceedings of the 2021 Conference of the North
  American Chapter of the Association for Computational Linguistics: Human
  Language Technologies}}. \bibinfo{publisher}{Association for Computational
  Linguistics}, \bibinfo{address}{Online}, \bibinfo{pages}{3030--3042}.
\newblock
\urldef\tempurl%
\url{https://doi.org/10.18653/v1/2021.naacl-main.241}
\showDOI{\tempurl}


\bibitem[\protect\citeauthoryear{Guu, Lee, Tung, Pasupat, and Chang}{Guu
  et~al\mbox{.}}{2020}]%
        {guu2020realm}
\bibfield{author}{\bibinfo{person}{Kelvin Guu}, \bibinfo{person}{Kenton Lee},
  \bibinfo{person}{Zora Tung}, \bibinfo{person}{Panupong Pasupat}, {and}
  \bibinfo{person}{Ming-Wei Chang}.} \bibinfo{year}{2020}\natexlab{}.
\newblock \bibinfo{title}{REALM: Retrieval-Augmented Language Model
  Pre-Training}.
\newblock
\newblock
\showeprint[arxiv]{2002.08909}~[cs.CL]


\bibitem[\protect\citeauthoryear{Hofst{\"a}tter, Lin, Yang, Lin, and
  Hanbury}{Hofst{\"a}tter et~al\mbox{.}}{2021}]%
        {Hofstaetter2021_tasb_dense_retrieval}
\bibfield{author}{\bibinfo{person}{Sebastian Hofst{\"a}tter},
  \bibinfo{person}{Sheng-Chieh Lin}, \bibinfo{person}{Jheng-Hong Yang},
  \bibinfo{person}{Jimmy Lin}, {and} \bibinfo{person}{Allan Hanbury}.}
  \bibinfo{year}{2021}\natexlab{}.
\newblock \showarticletitle{{Efficiently Teaching an Effective Dense Retriever
  with Balanced Topic Aware Sampling}}. In \bibinfo{booktitle}{\emph{Proc. of
  SIGIR}}.
\newblock


\bibitem[\protect\citeauthoryear{Hofstätter, Althammer, Schröder, Sertkan,
  and Hanbury}{Hofstätter et~al\mbox{.}}{2020}]%
        {hofstatter2020improving}
\bibfield{author}{\bibinfo{person}{Sebastian Hofstätter},
  \bibinfo{person}{Sophia Althammer}, \bibinfo{person}{Michael Schröder},
  \bibinfo{person}{Mete Sertkan}, {and} \bibinfo{person}{Allan Hanbury}.}
  \bibinfo{year}{2020}\natexlab{}.
\newblock \bibinfo{title}{Improving Efficient Neural Ranking Models with
  Cross-Architecture Knowledge Distillation}.
\newblock
\newblock
\showeprint[arxiv]{2010.02666}~[cs.IR]


\bibitem[\protect\citeauthoryear{Karpukhin, Oğuz, Min, Lewis, Wu, Edunov,
  Chen, and tau Yih}{Karpukhin et~al\mbox{.}}{2020}]%
        {karpukhin2020dense}
\bibfield{author}{\bibinfo{person}{Vladimir Karpukhin}, \bibinfo{person}{Barlas
  Oğuz}, \bibinfo{person}{Sewon Min}, \bibinfo{person}{Patrick Lewis},
  \bibinfo{person}{Ledell Wu}, \bibinfo{person}{Sergey Edunov},
  \bibinfo{person}{Danqi Chen}, {and} \bibinfo{person}{Wen tau Yih}.}
  \bibinfo{year}{2020}\natexlab{}.
\newblock \bibinfo{title}{Dense Passage Retrieval for Open-Domain Question
  Answering}.
\newblock
\newblock
\showeprint[arxiv]{2004.04906}~[cs.CL]


\bibitem[\protect\citeauthoryear{Khattab and Zaharia}{Khattab and
  Zaharia}{2020}]%
        {colbert}
\bibfield{author}{\bibinfo{person}{Omar Khattab} {and} \bibinfo{person}{Matei
  Zaharia}.} \bibinfo{year}{2020}\natexlab{}.
\newblock \showarticletitle{ColBERT: Efficient and Effective Passage Search via
  Contextualized Late Interaction over BERT}. In
  \bibinfo{booktitle}{\emph{Proceedings of the 43rd International ACM SIGIR
  Conference on Research and Development in Information Retrieval}} (Virtual
  Event, China) \emph{(\bibinfo{series}{SIGIR '20})}.
  \bibinfo{publisher}{Association for Computing Machinery},
  \bibinfo{address}{New York, NY, USA}, \bibinfo{pages}{39–48}.
\newblock
\showISBNx{9781450380164}
\urldef\tempurl%
\url{https://doi.org/10.1145/3397271.3401075}
\showDOI{\tempurl}


\bibitem[\protect\citeauthoryear{Lam, Pitrou, and Seibert}{Lam
  et~al\mbox{.}}{2015}]%
        {lam2015numba}
\bibfield{author}{\bibinfo{person}{Siu~Kwan Lam}, \bibinfo{person}{Antoine
  Pitrou}, {and} \bibinfo{person}{Stanley Seibert}.}
  \bibinfo{year}{2015}\natexlab{}.
\newblock \showarticletitle{Numba: A llvm-based python jit compiler}. In
  \bibinfo{booktitle}{\emph{Proceedings of the Second Workshop on the LLVM
  Compiler Infrastructure in HPC}}. \bibinfo{pages}{1--6}.
\newblock


\bibitem[\protect\citeauthoryear{Lin, Yang, and Lin}{Lin et~al\mbox{.}}{2021}]%
        {lin-etal-2021-batch}
\bibfield{author}{\bibinfo{person}{Sheng-Chieh Lin},
  \bibinfo{person}{Jheng-Hong Yang}, {and} \bibinfo{person}{Jimmy Lin}.}
  \bibinfo{year}{2021}\natexlab{}.
\newblock \showarticletitle{In-Batch Negatives for Knowledge Distillation with
  Tightly-Coupled Teachers for Dense Retrieval}. In
  \bibinfo{booktitle}{\emph{Proceedings of the 6th Workshop on Representation
  Learning for NLP (RepL4NLP-2021)}}. \bibinfo{publisher}{Association for
  Computational Linguistics}, \bibinfo{address}{Online},
  \bibinfo{pages}{163--173}.
\newblock
\urldef\tempurl%
\url{https://doi.org/10.18653/v1/2021.repl4nlp-1.17}
\showDOI{\tempurl}


\bibitem[\protect\citeauthoryear{MacAvaney, Nardini, Perego, Tonellotto,
  Goharian, and Frieder}{MacAvaney et~al\mbox{.}}{2020}]%
        {MacAvaney_2020}
\bibfield{author}{\bibinfo{person}{Sean MacAvaney},
  \bibinfo{person}{Franco~Maria Nardini}, \bibinfo{person}{Raffaele Perego},
  \bibinfo{person}{Nicola Tonellotto}, \bibinfo{person}{Nazli Goharian}, {and}
  \bibinfo{person}{Ophir Frieder}.} \bibinfo{year}{2020}\natexlab{}.
\newblock \showarticletitle{Expansion via Prediction of Importance with
  Contextualization}.
\newblock \bibinfo{journal}{\emph{Proceedings of the 43rd International ACM
  SIGIR Conference on Research and Development in Information Retrieval}}
  (\bibinfo{date}{Jul} \bibinfo{year}{2020}).
\newblock
\showISBNx{9781450380164}
\urldef\tempurl%
\url{https://doi.org/10.1145/3397271.3401262}
\showDOI{\tempurl}


\bibitem[\protect\citeauthoryear{Mallia, Khattab, Suel, and Tonellotto}{Mallia
  et~al\mbox{.}}{2021}]%
        {10.1145/3404835.3463030}
\bibfield{author}{\bibinfo{person}{Antonio Mallia}, \bibinfo{person}{Omar
  Khattab}, \bibinfo{person}{Torsten Suel}, {and} \bibinfo{person}{Nicola
  Tonellotto}.} \bibinfo{year}{2021}\natexlab{}.
\newblock \showarticletitle{Learning Passage Impacts for Inverted Indexes}. In
  \bibinfo{booktitle}{\emph{Proceedings of the 44th International ACM SIGIR
  Conference on Research and Development in Information Retrieval}} (Virtual
  Event, Canada) \emph{(\bibinfo{series}{SIGIR '21})}.
  \bibinfo{publisher}{Association for Computing Machinery},
  \bibinfo{address}{New York, NY, USA}, \bibinfo{pages}{1723–1727}.
\newblock
\showISBNx{9781450380379}
\urldef\tempurl%
\url{https://doi.org/10.1145/3404835.3463030}
\showDOI{\tempurl}


\bibitem[\protect\citeauthoryear{Nogueira and Cho}{Nogueira and Cho}{2019}]%
        {passage_ranking}
\bibfield{author}{\bibinfo{person}{Rodrigo Nogueira} {and}
  \bibinfo{person}{Kyunghyun Cho}.} \bibinfo{year}{2019}\natexlab{}.
\newblock \bibinfo{title}{Passage Re-ranking with BERT}.
\newblock
\newblock
\showeprint[arxiv]{1901.04085}~[cs.IR]


\bibitem[\protect\citeauthoryear{Nogueira and Lin}{Nogueira and Lin}{2019}]%
        {doct5}
\bibfield{author}{\bibinfo{person}{Rodrigo Nogueira} {and}
  \bibinfo{person}{Jimmy Lin}.} \bibinfo{year}{2019}\natexlab{}.
\newblock \bibinfo{title}{From doc2query to docTTTTTquery}.
\newblock
\newblock


\bibitem[\protect\citeauthoryear{Nogueira, Yang, Lin, and Cho}{Nogueira
  et~al\mbox{.}}{2019}]%
        {nogueira2019document}
\bibfield{author}{\bibinfo{person}{Rodrigo Nogueira}, \bibinfo{person}{Wei
  Yang}, \bibinfo{person}{Jimmy Lin}, {and} \bibinfo{person}{Kyunghyun Cho}.}
  \bibinfo{year}{2019}\natexlab{}.
\newblock \bibinfo{title}{Document Expansion by Query Prediction}.
\newblock
\newblock
\showeprint[arxiv]{1904.08375}~[cs.IR]


\bibitem[\protect\citeauthoryear{Paria, Yeh, Yen, Xu, Ravikumar, and
  Póczos}{Paria et~al\mbox{.}}{2020}]%
        {paria2020minimizing}
\bibfield{author}{\bibinfo{person}{Biswajit Paria}, \bibinfo{person}{Chih-Kuan
  Yeh}, \bibinfo{person}{Ian E.~H. Yen}, \bibinfo{person}{Ning Xu},
  \bibinfo{person}{Pradeep Ravikumar}, {and} \bibinfo{person}{Barnabás
  Póczos}.} \bibinfo{year}{2020}\natexlab{}.
\newblock \bibinfo{title}{Minimizing FLOPs to Learn Efficient Sparse
  Representations}.
\newblock
\newblock
\showeprint[arxiv]{2004.05665}~[cs.LG]


\bibitem[\protect\citeauthoryear{Paszke, Gross, Massa, Lerer, Bradbury, Chanan,
  Killeen, Lin, Gimelshein, Antiga, et~al\mbox{.}}{Paszke
  et~al\mbox{.}}{2019}]%
        {paszke2019pytorch}
\bibfield{author}{\bibinfo{person}{Adam Paszke}, \bibinfo{person}{Sam Gross},
  \bibinfo{person}{Francisco Massa}, \bibinfo{person}{Adam Lerer},
  \bibinfo{person}{James Bradbury}, \bibinfo{person}{Gregory Chanan},
  \bibinfo{person}{Trevor Killeen}, \bibinfo{person}{Zeming Lin},
  \bibinfo{person}{Natalia Gimelshein}, \bibinfo{person}{Luca Antiga},
  {et~al\mbox{.}}} \bibinfo{year}{2019}\natexlab{}.
\newblock \showarticletitle{PyTorch: An Imperative Style, High-Performance Deep
  Learning Library.}. In \bibinfo{booktitle}{\emph{NeurIPS}}.
\newblock


\bibitem[\protect\citeauthoryear{Qu, Ding, Liu, Liu, Ren, Zhao, Dong, Wu, and
  Wang}{Qu et~al\mbox{.}}{2021}]%
        {qu-etal-2021-rocketqa}
\bibfield{author}{\bibinfo{person}{Yingqi Qu}, \bibinfo{person}{Yuchen Ding},
  \bibinfo{person}{Jing Liu}, \bibinfo{person}{Kai Liu},
  \bibinfo{person}{Ruiyang Ren}, \bibinfo{person}{Wayne~Xin Zhao},
  \bibinfo{person}{Daxiang Dong}, \bibinfo{person}{Hua Wu}, {and}
  \bibinfo{person}{Haifeng Wang}.} \bibinfo{year}{2021}\natexlab{}.
\newblock \showarticletitle{{R}ocket{QA}: An Optimized Training Approach to
  Dense Passage Retrieval for Open-Domain Question Answering}. In
  \bibinfo{booktitle}{\emph{Proceedings of the 2021 Conference of the North
  American Chapter of the Association for Computational Linguistics: Human
  Language Technologies}}. \bibinfo{publisher}{Association for Computational
  Linguistics}, \bibinfo{address}{Online}, \bibinfo{pages}{5835--5847}.
\newblock
\urldef\tempurl%
\url{https://doi.org/10.18653/v1/2021.naacl-main.466}
\showDOI{\tempurl}


\bibitem[\protect\citeauthoryear{Reimers and Gurevych}{Reimers and
  Gurevych}{2019}]%
        {sentence_bert}
\bibfield{author}{\bibinfo{person}{Nils Reimers} {and} \bibinfo{person}{Iryna
  Gurevych}.} \bibinfo{year}{2019}\natexlab{}.
\newblock \showarticletitle{Sentence-BERT: Sentence Embeddings using Siamese
  BERT-Networks}. In \bibinfo{booktitle}{\emph{Proceedings of the 2019
  Conference on Empirical Methods in Natural Language Processing}}.
  \bibinfo{publisher}{Association for Computational Linguistics}.
\newblock
\urldef\tempurl%
\url{http://arxiv.org/abs/1908.10084}
\showURL{%
\tempurl}


\bibitem[\protect\citeauthoryear{Thakur, Reimers, R{\"{u}}ckl{\'{e}},
  Srivastava, and Gurevych}{Thakur et~al\mbox{.}}{2021}]%
        {beir_2021}
\bibfield{author}{\bibinfo{person}{Nandan Thakur}, \bibinfo{person}{Nils
  Reimers}, \bibinfo{person}{Andreas R{\"{u}}ckl{\'{e}}},
  \bibinfo{person}{Abhishek Srivastava}, {and} \bibinfo{person}{Iryna
  Gurevych}.} \bibinfo{year}{2021}\natexlab{}.
\newblock \showarticletitle{{BEIR:} {A} Heterogenous Benchmark for Zero-shot
  Evaluation of Information Retrieval Models}.
\newblock \bibinfo{journal}{\emph{CoRR}}  \bibinfo{volume}{abs/2104.08663}
  (\bibinfo{year}{2021}).
\newblock
\showeprint[arxiv]{2104.08663}
\urldef\tempurl%
\url{https://arxiv.org/abs/2104.08663}
\showURL{%
\tempurl}


\bibitem[\protect\citeauthoryear{Tu, Yang, Fu, Xie, Tan, Xiong, Li, and Lin}{Tu
  et~al\mbox{.}}{2020}]%
        {tu2020approximate}
\bibfield{author}{\bibinfo{person}{Zhengkai Tu}, \bibinfo{person}{Wei Yang},
  \bibinfo{person}{Zihang Fu}, \bibinfo{person}{Yuqing Xie},
  \bibinfo{person}{Luchen Tan}, \bibinfo{person}{Kun Xiong},
  \bibinfo{person}{Ming Li}, {and} \bibinfo{person}{Jimmy Lin}.}
  \bibinfo{year}{2020}\natexlab{}.
\newblock \showarticletitle{Approximate Nearest Neighbor Search and Lightweight
  Dense Vector Reranking in Multi-Stage Retrieval Architectures}. In
  \bibinfo{booktitle}{\emph{Proceedings of the 2020 ACM SIGIR on International
  Conference on Theory of Information Retrieval}}. \bibinfo{pages}{97--100}.
\newblock


\bibitem[\protect\citeauthoryear{Wolf, Debut, Sanh, Chaumond, Delangue, Moi,
  Cistac, Rault, Louf, Funtowicz, Davison, Shleifer, von Platen, Ma, Jernite,
  Plu, Xu, Scao, Gugger, Drame, Lhoest, and Rush}{Wolf et~al\mbox{.}}{2020}]%
        {wolf2020huggingfaces}
\bibfield{author}{\bibinfo{person}{Thomas Wolf}, \bibinfo{person}{Lysandre
  Debut}, \bibinfo{person}{Victor Sanh}, \bibinfo{person}{Julien Chaumond},
  \bibinfo{person}{Clement Delangue}, \bibinfo{person}{Anthony Moi},
  \bibinfo{person}{Pierric Cistac}, \bibinfo{person}{Tim Rault},
  \bibinfo{person}{Rémi Louf}, \bibinfo{person}{Morgan Funtowicz},
  \bibinfo{person}{Joe Davison}, \bibinfo{person}{Sam Shleifer},
  \bibinfo{person}{Patrick von Platen}, \bibinfo{person}{Clara Ma},
  \bibinfo{person}{Yacine Jernite}, \bibinfo{person}{Julien Plu},
  \bibinfo{person}{Canwen Xu}, \bibinfo{person}{Teven~Le Scao},
  \bibinfo{person}{Sylvain Gugger}, \bibinfo{person}{Mariama Drame},
  \bibinfo{person}{Quentin Lhoest}, {and} \bibinfo{person}{Alexander~M. Rush}.}
  \bibinfo{year}{2020}\natexlab{}.
\newblock \bibinfo{title}{HuggingFace's Transformers: State-of-the-art Natural
  Language Processing}.
\newblock
\newblock
\showeprint[arxiv]{1910.03771}~[cs.CL]


\bibitem[\protect\citeauthoryear{Xiong, Xiong, Li, Tang, Liu, Bennett, Ahmed,
  and Overwijk}{Xiong et~al\mbox{.}}{2020}]%
        {xiong2020approximate}
\bibfield{author}{\bibinfo{person}{Lee Xiong}, \bibinfo{person}{Chenyan Xiong},
  \bibinfo{person}{Ye Li}, \bibinfo{person}{Kwok-Fung Tang},
  \bibinfo{person}{Jialin Liu}, \bibinfo{person}{Paul Bennett},
  \bibinfo{person}{Junaid Ahmed}, {and} \bibinfo{person}{Arnold Overwijk}.}
  \bibinfo{year}{2020}\natexlab{}.
\newblock \bibinfo{title}{Approximate Nearest Neighbor Negative Contrastive
  Learning for Dense Text Retrieval}.
\newblock
\newblock
\showeprint[arxiv]{2007.00808}~[cs.IR]


\bibitem[\protect\citeauthoryear{Xiong, Xiong, Li, Tang, Liu, Bennett, Ahmed,
  and Overwikj}{Xiong et~al\mbox{.}}{2021}]%
        {xiong2021approximate}
\bibfield{author}{\bibinfo{person}{Lee Xiong}, \bibinfo{person}{Chenyan Xiong},
  \bibinfo{person}{Ye Li}, \bibinfo{person}{Kwok-Fung Tang},
  \bibinfo{person}{Jialin Liu}, \bibinfo{person}{Paul~N. Bennett},
  \bibinfo{person}{Junaid Ahmed}, {and} \bibinfo{person}{Arnold Overwikj}.}
  \bibinfo{year}{2021}\natexlab{}.
\newblock \showarticletitle{Approximate Nearest Neighbor Negative Contrastive
  Learning for Dense Text Retrieval}. In
  \bibinfo{booktitle}{\emph{International Conference on Learning
  Representations}}.
\newblock
\urldef\tempurl%
\url{https://openreview.net/forum?id=zeFrfgyZln}
\showURL{%
\tempurl}


\bibitem[\protect\citeauthoryear{Yang, Lu, Yang, and Lin}{Yang
  et~al\mbox{.}}{2019}]%
        {neural_hype}
\bibfield{author}{\bibinfo{person}{Wei Yang}, \bibinfo{person}{Kuang Lu},
  \bibinfo{person}{Peilin Yang}, {and} \bibinfo{person}{Jimmy Lin}.}
  \bibinfo{year}{2019}\natexlab{}.
\newblock \showarticletitle{Critically Examining the “Neural Hype”}.
\newblock \bibinfo{journal}{\emph{Proceedings of the 42nd International ACM
  SIGIR Conference on Research and Development in Information Retrieval}}
  (\bibinfo{date}{Jul} \bibinfo{year}{2019}).
\newblock
\showISBNx{9781450361729}
\urldef\tempurl%
\url{https://doi.org/10.1145/3331184.3331340}
\showDOI{\tempurl}


\bibitem[\protect\citeauthoryear{Zamani, Dehghani, Croft, Learned-Miller, and
  Kamps}{Zamani et~al\mbox{.}}{2018}]%
        {snrm}
\bibfield{author}{\bibinfo{person}{Hamed Zamani}, \bibinfo{person}{Mostafa
  Dehghani}, \bibinfo{person}{W.~Bruce Croft}, \bibinfo{person}{Erik
  Learned-Miller}, {and} \bibinfo{person}{Jaap Kamps}.}
  \bibinfo{year}{2018}\natexlab{}.
\newblock \showarticletitle{From Neural Re-Ranking to Neural Ranking: Learning
  a Sparse Representation for Inverted Indexing}. In
  \bibinfo{booktitle}{\emph{Proceedings of the 27th ACM International
  Conference on Information and Knowledge Management}} (Torino, Italy)
  \emph{(\bibinfo{series}{CIKM '18})}. \bibinfo{publisher}{Association for
  Computing Machinery}, \bibinfo{address}{New York, NY, USA},
  \bibinfo{pages}{497–506}.
\newblock
\showISBNx{9781450360142}
\urldef\tempurl%
\url{https://doi.org/10.1145/3269206.3271800}
\showDOI{\tempurl}


\bibitem[\protect\citeauthoryear{Zhao, Lu, and Lee}{Zhao et~al\mbox{.}}{2020}]%
        {zhao2020sparta}
\bibfield{author}{\bibinfo{person}{Tiancheng Zhao}, \bibinfo{person}{Xiaopeng
  Lu}, {and} \bibinfo{person}{Kyusong Lee}.} \bibinfo{year}{2020}\natexlab{}.
\newblock \bibinfo{title}{SPARTA: Efficient Open-Domain Question Answering via
  Sparse Transformer Matching Retrieval}.
\newblock
\newblock
\showeprint[arxiv]{2009.13013}~[cs.CL]


\end{thebibliography}

\end{document}